\documentstyle[epsf,12pt]{ioplppt}

\begin{document}
\JPA

\title{Bethe Ansatz solution of a decagonal rectangle triangle random
  tiling.}[Solution of a decagonal random tiling]

\author{Jan de Gier and Bernard Nienhuis}
\address{Instituut voor Theoretische Fysica, Universiteit van Amsterdam,
Valckenierstraat~65, 1018 XE Amsterdam, The
Netherlands.\ftnote{1}{Electronic mail addresses: {\tt
    degier@phys.uva.nl} and {\tt nienhuis@phys.uva.nl}}}
\ftnote{0}{FAX: +(31) 20 525 5778}
\begin{abstract}
A random tiling of rectangles and triangles displaying a decagonal
phase is solved by Bethe Ansatz. Analogously to the solutions of the
dodecagonal square triangle and the octagonal rectangle triangle
tiling an exact expression for the maximum of the entropy is found. 
\end{abstract}

\maketitle

\section{Introduction}
The discussion on the stability of quasi-crystals has not resulted in a
general consensus yet. Even very recently, arguments against
\cite{Coddens:1997} and in favour of \cite{Joseph:1997a} the random
tiling scenario have appeared in the literature. It has also been
suggested that the entropically stabilised state results from
quasi-crystal growth \cite{Joseph:1997b}. From the point of view of
statistical mechanics random tiling models are very interesting, not
in the least because some of them allow for an exact solution. This
means that one is able to derive exact expressions for the entropy
and other thermodynamic quantities. In this paper we present a random
tiling with a decagonal phase which is solvable by the Bethe Ansatz
method, very much in analogy with the dodecagonal square-triangle and
the octagonal rectangle-triangle tilings
\cite{Widom:1993,Kalugin:1994,Gier:1996a,Gier:1997b}. The random
tiling with a decagonal phase is the ensemble of tilings of the plane
by rectangles and isosceles triangles with sides of length 1 and
$l=2\sin (\pi/5)=\sqrt{2+\tau}/\tau$, where $\tau=(\sqrt{5}+1)/2$ is
the golden mean. Cockayne \cite{Cockayne:1995} devised an inflation
rule for this set of tiles that constructs a tiling with decagonal
symmetry. This tiling corresponds to a maximally dense decagonal disc
packing under the condition that nearest neighbour vectors are limited
to certain directions. The {\em random} tiling with additional
constraints has been studied by Oxborrow and Mihalkovi\v c
\cite{Oxborrow:1995} to model d-AlPdMn. More recently, Roth and
Henley \cite{Roth:1997} used the rectangle-triangle random tiling and
some of its sub-ensembles to model decagonal quasi-crystal structures
resulting from a molecular dynamics simulation. 

\begin{figure}[h]
\centerline{
\begin{picture}(330,180)
\put(0,5){\epsffile{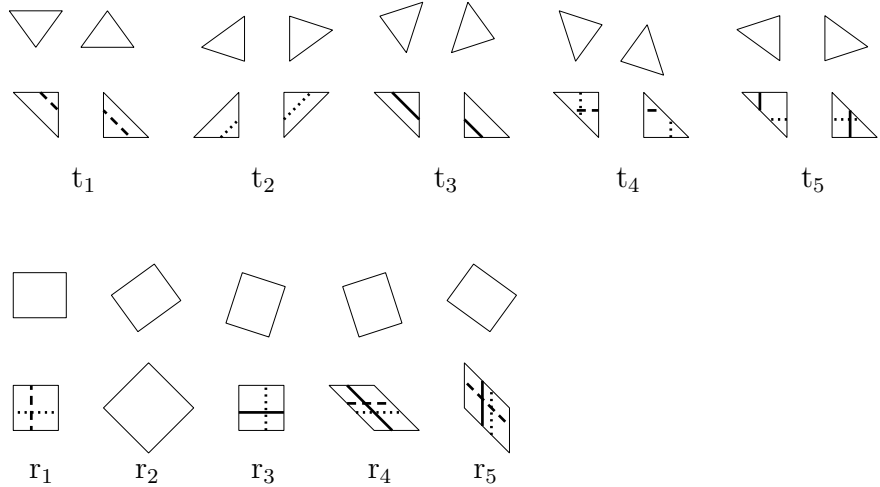}}
\put(8,0){${\rm r}_1$}
\put(48,0){${\rm r}_2$}
\put(92,0){${\rm r}_3$}
\put(136,0){${\rm r}_4$}
\put(176,0){${\rm r}_5$}
\put(24,110){${\rm t}_1$}
\put(92,110){${\rm t}_2$}
\put(161,110){${\rm t}_3$}
\put(230,110){${\rm t}_4$}
\put(300,110){${\rm t}_5$}
\end{picture}}
\caption{Tiles of the the tenfold tiling with rectangles and triangles and
their deformations to the regular square lattice. The five different
orientations of the pairs of triangles and those of the rectangles are
denoted by ${\rm t}_i$ and ${\rm r}_i$ respectively.}
\label{fig:tiledef}
\end{figure}

One of the main goals in statistical mechanics is the calculation of
the partition sum which, for random tilings, is the weighted sum over
all possible tiling configurations. To be able to enumerate all
possible tilings we use the transfer matrix method, for which it is
convenient to transform the tiling to a model on the square
lattice. This transformation is depicted in \fref{fig:tiledef}. The
triangles always come in pairs which are denoted by ${\rm t}_1,
\ldots, {\rm t}_5$ corresponding to their five different
orientations. Similarly the five different orientations of the
rectangles are labeled by ${\rm r}_i$. The decorations on the deformed
tiles are such that continuity of the decorating lines is equivalent
to the restriction that the tiles fit together without gaps or
overlaps. In this way every decorated tiling of the lattice with the
deformed tiles corresponds to an allowed tiling of the plane by the
original rectangles and triangles. Using this transformation every
vertex of the tiling falls on a vertex of the square lattice, though
the reverse is not true.  The partition sum of the deformed model on
the lattice is now defined by  
\begin{equation}
Z = \sum_{\cal C}\, \prod_{i=1}^5 r_i^{N_{{\rm r}_i} ({\cal C})}
t_i^{N_{{\rm t}_i} ({\cal C})},\label{eq:Zlat}
\end{equation}
where the sum is over all possible configurations ${\cal C}$. For a
given configuration ${\cal C}$, we denote the number of deformed
rectangles ${\rm r}_i$ by $N_{{\rm r}_i} ({\cal C})$ and the number of
deformed triangles ${\rm t}_i$ by $N_{{\rm t}_i} ({\cal C})$. The
partition sum is thus a weighted sum where each deformed triangle ${\rm
  t}_i$ and each deformed rectangle ${\rm r}_i$ is assigned a weight
$t_i$ and $r_i$ respectively. The partition sum \eref{eq:Zlat} is
equal the partition sum for the tiling, provided that one chooses the
weights of the deformed tiles properly \cite{Li:1992}. The latter is a
consequence of the transformation which changes the areas of the
various tiles differently.

The definition of the transfer matrix {\bf T} of the
deformed model is the obvious one on the square lattice. The
horizontal edges of the square lattice can be in one out of five
possible states, represented by the presence or absence of the
decorating lines. The transfer matrix elements $T_{ij}$ between two
sequences $i$ and $j$ of such horizontal edges is equal to the product
of the weights of the deformed tiles that fit in between $i$ and
$j$. As is well known, the partition sum per row in the thermodynamic
limit is given by the largest eigenvalue of the transfer matrix. The
free energy is then given by the logarithm of this largest eigenvalue. 

Recall that according to the random tiling hypotheses the entropy per
area may be written as \cite{Henley:1988} 
\begin{equation}
\sigma_{\rm a} = \sigma_{{\rm a},0} - \case{1}{4} K_1 I_1 -
\case{1}{4} K_2 I_2  - \case{1}{2} K_3 I_3,
\end{equation}
where the $I_j$ are the quadratic phason strain invariants for the
$3\times 2$ phason strain tensor. The goal of our work is to calculate
the residual entropy $\sigma_{{\rm a},0}$ and the phason strain
elastic constants $K_j$. In \sref{se:BA} we derive the Bethe Ansatz
(BA) equations that diagonalise the transfer matrix for this model and
which already give a huge reduction of the numerical problem. It turns
out that for this tiling the method of Kalugin
\cite{Kalugin:1994,Gier:1997b} is applicable to solve these BA
equations . This then enables us to calculate the residual entropy for
this tiling exactly. The calculation of the elastic constants poses
some problems and in this paper we will concentrate on the maximum 
only. It will be shown that the maximum of the entropy per vertex of
this decagonal random tiling model is given by   
\begin{equation}
\sigma_{\rm v} = \frac{1}{2} \left( \log \frac{5^5}{4^4} - 2\sqrt{5}
\log \tau \right). 
\end{equation}

\section{Degrees of freedom}
In principle there are fifteen partial densities for this tiling,
corresponding to the ten different orientations of the triangles and
the five different orientations of the rectangles. Since the triangles
always occur in pairs, we are left with ten degrees of freedom. One of
these is removed by the fact that the total area is constant. Furthermore,
there are three nonlinear geometrical constraints, so that the phase
space of this random tiling is six-dimensional. The geometrical
constraints are derived in the following.\footnote{This argument makes
  use of the periodic boundary conditions imposed by placing the
  lattice model on a cylinder. We believe that for free boundary
  conditions modified versions of such constraints hold, involving the
  configuration of the boundary.} The first thing to note is that the
triangles can be viewed as domain walls between patches consisting
solely of one type of rectangle. This phenomenon is similar to what
happens in de dodecagonal square-triangle and the octagonal
rectangle-triangle tilings. In our choice of decoration in
\fref{fig:tiledef} it is easily seen that between patches consisting
only of the tile r$_2$ there are three different types of domain wall,
denoted by the solid, dotted and dashed lines. Two types of domain
wall run from bottom right to top left. They are drawn as solid and
dashed lines and we denote their number by $n_1$ and  $n_2$
respectively. The other type of domain wall runs from bottom left to
top right and is drawn as a dotted line. Their number is given by
$m$. We denote the average number of triangles and rectangles per
layer by $n_{{\rm t}_i}$ and $n_{{\rm r}_i}$ respectively. It is then
easily seen from \fref{fig:tiledef} that  
\begin{equation}
\eqalign{
n_1 &= \case{1}{2}(n_{{\rm t}_3} + n_{{\rm t}_5}) + n_{{\rm r}_4} + n_{{\rm
    r}_5},\\
n_2 &= \case{1}{2} n_{{\rm t}_1} + n_{{\rm r}_1} + n_{{\rm r}_5},\\
m &= \case{1}{2} (n_{{\rm t}_2} + n_{{\rm t}_4}) + n_{{\rm r}_3} + n_{{\rm
    r}_5}.} \label{eq:part2tiles}
\end{equation}
Let $p_1$ be the number of layers such that each dotted line crosses each
dashed line once. The number of such crossings in this patch of size
$p_1N$ is then 
\begin{equation}
n_2 m = p_1(n_{{\rm r}_1} + n_{{\rm r}_5} + \case{1}{2}n_{{\rm t}_4}).
\end{equation}
Similarly, if $p_2$ is the number of layers such that each
dotted line crosses each solid line once, and $q$ the number of layers
such that each dashed line crosses each solid line once we find
\begin{equation}
\eqalign{
n_1 m &= p_2(n_{{\rm r}_3} + n_{{\rm r}_4} + \case{1}{2}n_{{\rm t}_5}),\\
n_1 n_2 &= q(n_{{\rm r}_4} + n_{{\rm r}_5}).}
\end{equation}
The numbers $p_1$, $p_2$ and $q$ can also be calculated in another
way. For this, it is convenient to introduce the average shift per
layer ($s_1$, $s_2$ and $s_m$) of each domain wall. These are given by
\begin{equation}
\eqalign{
s_1 &= -1+\frac{1}{n_1} (n_{{\rm r}_5} - n_{{\rm r}_3} + \case{1}{2}
n_{{\rm t}_5}),\\ 
s_2 &= -1+\frac{1}{n_2} (n_{{\rm r}_1} - n_{{\rm r}_4} - \case{1}{2}
n_{{\rm t}_4}),\\ 
s_m &= 1-\frac{1}{m} (n_{{\rm r}_3} + n_{{\rm r}_5} - n_{{\rm r}_1} -
n_{{\rm r}_4} + \case{1}{2} (n_{{\rm t}_4} - n_{{\rm t}_5})).}
\end{equation}
From the condition that each dotted line must cross each dashed line
once in a patch of size $p_1N$ it then follows that in $p_1$ layers
the average relative shift of the domain walls must be equal to
$N$. The same arguments applied to the other cases then gives
\begin{equation}
N = p_1 (s_m-s_2) = p_2 (s_m-s_1) = q (s_1-s_2).
\end{equation}
Putting all these equations together, using the fact that the system
size $N=n_1+n_2+m+2 (n_{{\rm r}_2} - n_{{\rm r}_5})$ one finds three
independent relations among the tile densities. These can be rewritten
as follows 
\begin{eqnarray}
n_{{\rm t}_1} (n_{{\rm t}_2}+n_{{\rm t}_5}) &=& 2 (n_{{\rm
    t}_4}+2(n_{{\rm r}_4}+n_{{\rm r}_5})) (n_{{\rm r}_1} + n_{{\rm
    r}_2})\nonumber \\ 
&& {}+ 2 (n_{{\rm t}_3}+2(n_{{\rm r}_3}+n_{{\rm r}_2}))
(n_{{\rm r}_1} + n_{{\rm r}_5}),\label{eq:Constr1}
\end{eqnarray}
\begin{eqnarray}
n_{{\rm t}_2} ( n_{{\rm t}_3} + 2(n_{{\rm r}_4}+n_{{\rm r}_5})) + 4
n_{{\rm r}_5} (n_{{\rm r}_3}+n_{{\rm r}_4}) = \nonumber\\
\hphantom{n_{{\rm t}_2} ( n_{{\rm t}_3} + 2(n_{{\rm r}_4}+n_{{\rm r}_5}))}
n_{{\rm t}_5} ( n_{{\rm t}_4} + 2(n_{{\rm r}_3}+n_{{\rm r}_2})) + 4
n_{{\rm r}_2} (n_{{\rm r}_3}+n_{{\rm r}_4}),\label{eq:Constr2}
\end{eqnarray}
which are symmetric under the simultaneous exchange of indices $2
\leftrightarrow 5$ and $3 \leftrightarrow 4$, i.e. the mirror symmetry in
the $y$-axis. Other relations may be obtained from
\eref{eq:Constr1} and \eref{eq:Constr2} by applying a
rotation over $2\pi/5$, i.e. shifting each index by one. Only three of
them, however, are independent.  

At the symmetric point each orientation occurs equally often, so we
have $n_{{\rm t}_i}= \case{1}{5} n_{\rm tri}$ and $n_{{\rm r}_i}=
\case{1}{5} n_{\rm rect}$, where $n_{\rm tri}$ and $n_{\rm rect}$ are
the average total numbers of triangles and rectangles per layer. It
follows then from \eref{eq:Constr1} and the expression for the system
size that at the symmetric point $n_{\rm tri} = 10 \tau^{-4}N$ and
$n_{\rm rect} = \case{5}{2} \tau^{-5}N$. Using \eref{eq:part2tiles}
we find that these values correspond with the following numbers of
domain walls,
\begin{equation}
\tau n_2 = n_1 = m = \tau^{-2}N. \label{eq:QCpartdens}
\end{equation} 

\section{Bethe Ansatz \label{se:BA}}
In this section we derive the Bethe Ansatz equations for the lattice
model to diagonalise the transfer matrix. Again we make use of the
fact that the triangles can be viewed as domain walls. Since
these domain walls persist through the lattice we can think of them as
being trajectories of three types of particles, where each number of
particles is conserved. The transfer matrix {\bf T}, acting in the
upwards direction, can be thought of as an evolution operator for
these particles. Two types of the particles are then left movers whose
trajectories are given by the solid and dashed lines. We call them of
type 1 and type 2 respectively. The other type is
a right mover and its trajectories are drawn as dotted lines. Because
the number of particles of each type is conserved, {\bf T} is block
diagonal in the particle numbers. In the sequel we shall diagonalise
{\bf T} in each block  separately by using a nested coordinate Bethe
Ansatz \cite{Baxter:1970}. 

A state $\alpha$ on a row of the lattice can be specified by the
positions $y_1,\dots,y_m$ of the right movers and by the positions
$x_1,\ldots,x_n$ of the left movers with the specification
that the lines $i_1,\ldots,i_{n_1}$ at positions
$x_{i_1},\ldots,x_{i_{n_1}}$ are of type 1. Elements $\psi(\alpha)$
of an eigenvector of {\bf T} thus can be written explicitly as
$\psi(i_1,\dots,i_{n_1}|x_1,\ldots,x_n;y_1,\ldots,y_m)$. The state
which has no particles, i.e. the one with only the rectangles r$_2$,
is called the pseudo-vacuum. We make the following Ansatz for
the form of the eigenvector,
\begin{eqnarray}
\psi(i_1,\dots,i_{n_1}|x_1,\ldots,x_n;y_1,\ldots,y_m)=\nonumber\\
\sum_{\pi,\rho} \sum_\mu A(\Gamma) B(\mu)
\prod_{a=1}^n z_{\pi_a}^{x_a} \prod_{b=1}^m w_{\rho_b}^{y_b}
\prod_{c=1}^{n_1} \left[ d^{x_{i_c}}\prod_{r=1}^{i_c-1}
u(\mu_c,\pi_r)\right], \label{eq:eigvec_ans} 
\end{eqnarray}
where the sum runs over all permutations $\mu=(\mu_1,\ldots,\mu_{n_1})$ of
the numbers $1,\ldots,n_1$, all permutations $\pi$ of the numbers
$1,\ldots,n$ and all permutations $\rho$ of the numbers
$1,\ldots,m$. The coordinates $x_a$ and $y_b$ enter the Ansatz
\eref{eq:eigvec_ans} as powers of complex numbers, $z_i$ and
$w_j$ respectively, that are to be determined later. Factors in the
eigenvector due to the order of the different types of right
movers are given by the expression between brackets in
\eref{eq:eigvec_ans}. This is the so called nested part of the Ansatz
which is a generalisation of the simple power for the coordinates. The
index $i_c$ may be seen as the `coordinate' of right movers of type 1
relative to all right movers. This coordinate $i_c$ determines the
upper limit of a product over a complex valued function $u$ which may
still depend on the numbers $z_i$. The Ansatz \eref{eq:eigvec_ans}
further contains an adjustable constant $d$ whose presence will
become clear below. 

The amplitudes $A$ depend on the permutations $\pi$ and
$\rho$ and on the configuration of the left and right movers. These
together are coded in a vector $\Gamma$ in the following way. Let {\bf
  p} be the vector of coordinates $x_i$ and $y_j$ of all domain walls,
ordered so that $p_j < p_{j+1}$. The entries of $\Gamma$ are the
elements of the permutations $\pi$ and $\rho$. The order of succession
in $\Gamma$ of elements taken from $\pi$ and $\rho$ matches that of
the elements of $x$ and $y$ respectively in {\bf p}. The amplitudes
$B$ only depend on the permutation $\mu$. 

\begin{figure}[h]
\centerline{\epsffile{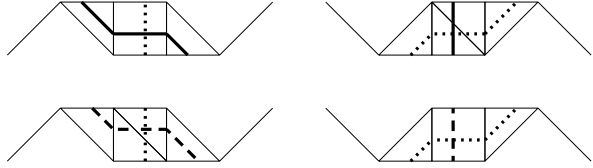}}
\caption{Two-particle collision diagrams. Upper line: Two types of
collision of walls of type 1 and 3. Second line: Two types of collision
of walls of type 2 and 3.}
\label{fig:2partcol}
\end{figure}

If all the domain walls are separated the action of the transfer
matrix is just a shift of each line to the right or to the left.
The eigenvalue of {\bf T} corresponding to the vector
\eref{eq:eigvec_ans} is therefore given by
\begin{equation}
\Lambda \,=\, r_2^{(N-n-m)/2}  t_1^{2n-2n_1} t_2^{2m} (t_3^2 d)^{n_1}\,
\prod_{a=1}^n z_a \prod_{b=1}^m w_b^{-1},\label{eigval}
\end{equation}
where $N$ is the size of the lattice.
At places where different domain walls are close together, the action
of {\bf T} is not given by a mere shift of all domain walls. Whenever
there is a right mover just in front of a left 
mover and there are no other neighbouring walls, two things can
happen. Either the right mover jumps over the left mover, which does
not move, or the left mover jumps over the right mover, see
\fref{fig:2partcol}. 
These exceptions imply the following relations for
\eref{eq:eigvec_ans} to be an eigenvector of {\bf T}.
\begin{equation}
\frac{A(\ldots,\pi_k,\rho_l,\ldots)}{A(\ldots,\rho_l,\pi_k,\ldots)} =
\frac{r_3 d}{t_2^2} z_{\pi_k} w_{\rho_l} + \frac{r_1}{t_1^2}
z_{\pi_k}^{-1} w_{\rho_l}^{-1},\label{eq:rlscat},
\end{equation}
and they put the following constraints on the weights
\begin{equation}
r_3 d = t_4^2 \text{~~and~~} r_3 t_1^2 t_5^2 = r_1 t_3^2 t_4^2.  
\end{equation}
The reason for the constant $d$ in the Ansatz \eref{eq:eigvec_ans} is
now clear: if it were omitted there would be an additional constraint
between the weights $r_3$ and $t_4$. Scattering processes involving
three particles of only two different types give the following
restrictions on the amplitudes $A$,
\begin{equation}
\frac{A(\ldots,\pi_k,\pi_{k+1},\ldots)}{A(\ldots,\pi_{k+1},\pi_k,\ldots)}
= -\frac{z_{\pi_{k+1}}}{z_{\pi_k}},\quad
\frac{A(\ldots,\rho_l,\rho_{l+1},\ldots)}{A(\ldots,\rho_{l+1},\rho_l,\ldots)}
= -\frac{w_{\rho_{l+1}}}{w_{\rho_l}}.\label{eq:rrscat}
\end{equation}
From these processes it also follows that the amplitudes $B$ obey the
equation 
\begin{equation}
\sum_{\mu_p,\mu_{p+1}} B(\ldots,\mu_p,\mu_{p+1},\ldots) z_{\pi_k}
(u(\mu_{p+1},\pi_k)-u(\mu_{p+1},\pi_{k+1}))=0. \label{eq:Bconstr}
\end{equation}
By the sum in \eref{eq:Bconstr} we mean a sum over both permutations
of the numbers $\mu_p$ and $\mu_{p+1}$. A similar notation is used in
the next two equations, which result from scattering processes with
three particles which are all of a different type. 
\begin{eqnarray}
\sum_{\pi_k,\pi_{k+1}} A(\ldots,\pi_k,\pi_{k+1},\ldots) z_{\pi_k}
u(\mu_p,\pi_k) =\nonumber\\
\sum_{\pi_k,\pi_{k+1}} A(\ldots,\pi_k,\pi_{k+1},\ldots)\left(
\frac{r_2 r_3 r_5}{t_1^2 t_3^2 t_4^4} z_{\pi_k}^{-1} + \frac{t_1^2
  r_4}{t_3^2 r_1} z_{\pi_k} z_{\pi_{k+1}}^2\right). 
\label{eq:Aconstr}
\end{eqnarray}
Equation \eref{eq:Aconstr} is satisfied (after $A$ is eliminated using
\eref{eq:rrscat}), if $u$ is of the form
\begin{equation}
u(\mu_p,\pi_k) = v_{\mu_p} + \frac{r_2 r_3 r_5}{t_3^2 t_1^2 t_4^4}
z_{\pi_k}^{-2} - \frac{t_1^2 r_4}{t_3^2 r_1} z_{\pi_k}^2,
\label{eq:usol}
\end{equation}
with any complex number $v_{\mu_p}$. Substituting \eref{eq:usol} into
\eref{eq:Bconstr} it follows that the amplitudes $B$ fulfil the relation 
\begin{equation}
\frac{B(\ldots,\mu_p,\mu_{p+1},\ldots)}{B(\ldots,\mu_{p+1},\mu_p,\ldots)}
= -1.\label{eq:r1r1scat}
\end{equation}
Using periodic boundary conditions, it follows from the form of the
eigenvector \eref{eq:eigvec_ans} and the relations \eref{eq:rlscat},
\eref{eq:rrscat} and \eref{eq:r1r1scat} that the
complex numbers $z_i$, $w_j$ and $v_k$ should obey the following
Bethe Ansatz equations
\begin{eqnarray}
w_j^{-N} &=& (-)^{m-1} \prod_{k=1}^m \left( \frac{w_j}{w_k}\right)
\prod_{i=1}^n \left( \frac{t_4^2}{t_2^2} z_i w_j + \frac{r_1}{t_1^2}
z_i^{-1} w_j^{-1}\right),\\
z_i^N &=& (-)^{n-1}\prod_{k=1}^n \left( \frac{z_k}{z_i}\right)
\prod_{j=1}^m \left( \frac{t_4^2}{t_2^2} z_i w_j + \frac{r_1}{t_1^2}
z_i^{-1} w_j^{-1}\right)\times\nonumber\\
&& \prod_{l=1}^{n_1} \left(v_l +
\frac{r_2 r_3 r_5}{t_1^2 t_3^2 t_4^4} z_i^{-2} - \frac{t_1^2 r_4}{t_3^2
  r_1} z_i^2\right),\\
(-)^{n_1-1} &=& \left( \frac{t_4^2}{r_3} \right)^N\prod_{i=1}^n
\left(v_l + \frac{r_2 r_3 r_5}{t_1^2 t_3^2 t_4^4} z_i^{-2} -
  \frac{t_1^2 r_4}{t_3^2 r_1} z_i^2\right).
\end{eqnarray}
To rewrite the BA equations in a more suitable form we introduce the
following variables,
\begin{equation}
\eqalign{
&\tilde{\xi}_i = \left(\frac{r_4 t_1^4 t_4^4}{r_1 r_2 r_3
  r_5}\right)^{1/2} z_i^2, \quad 
\psi_j = - \left(\frac{r_1 r_4 t_2^4}{r_2 r_3 r_5}\right)^{1/2}
w_j^{-2},\\
&u_l - u_l^{-1} = \left(\frac{r_1 t_3^4 t_4^4}{r_2 r_3 r_4
  r_5}\right)^{1/2} v_l.} 
\end{equation}
The BA equations then become
\begin{eqnarray}
(-\psi_j)^{(N+n+m)/2} = (-)^{m-1} C \prod_{i=1}^n
\tilde{\xi}_i^{-1/2} \prod_{k=1}^m (-\psi_k)^{1/2} \prod_{i=1}^n
(\tilde{\xi}_i-\psi_j),  
\label{eq:BAEpsi1}\\
\tilde{\xi}_i^{(N+n+m)/2} = (-)^{n-1} D \prod_{k=1}^n
\tilde{\xi}_k^{1/2} \prod_{j=1}^m (-\psi_j)^{-1/2} \prod_{l=1}^{n_1} u_l^{-1}
\times\nonumber\\  
\hphantom{\tilde{\xi}_i^{(N+n+m)/2} = }{} \prod_{j=1}^m (\tilde{\xi}_i-\psi_j)
\prod_{l=1}^{n_1} (u_l-\tilde{\xi}_i)(u_l+\tilde{\xi}_i^{-1}),
\label{eq:BAEksi1}\\  
u_l^{n} = (-)^{n_1-1} E \prod_{i=1}^n
(u_l-\tilde{\xi}_i)(u_l+\tilde{\xi}_i^{-1}), \label{eq:BAEu1} 
\end{eqnarray} 
where $C,D$ and $E$ are given by
\begin{equation}
\eqalign{
C =& \left( \frac{r_1 r_4 t_2^4}{r_2 r_3 r_5} \right)^{N/4} \left(
\frac{r_1 t_4^2}{t_1^2 t_2^2} \right)^{n/2} \\
D =&\; \left( \frac{r_4 t_1^4 t_4^4}{r_1 r_2 r_3 r_5} \right)^{N/4}
\left( \frac{r_1 t_4^2}{t_1^2 t_2^2} \right)^{m/2} \left( \frac{r_2
    r_3 r_4 r_5}{r_1 t_3^4 t_4^4} \right)^{n_1/2}\\
E =& \left( \frac{t_4^2}{r_3} \right)^N \left( \frac{r_2 r_3 r_4
  r_5}{r_1 t_3^4 t_4^4} \right)^{n/2}.} \label{eq:CDEdef}
\end{equation}  
The eigenvalue in terms of these new variables is then given by 
\begin{equation}
\Lambda = r_2^{N/2} \left( \frac{r_1 r_3 r_5 t_1^4}{r_2 r_4 t_4^4}
\right)^{n/4} \left( \frac{r_3 r_5 t_2^4}{r_1 r_2 r_4} \right)^{m/4}
\left( \frac{t_3^2 t_4^2}{r_3 t_1^2} \right)^{n_1} \prod_{i=1}^n
  \tilde{\xi}_i^{1/2} \prod_{j=1}^m (-\psi_j)^{1/2}. \label{eq:eigval}
\end{equation}

In summary we have shown in this section that the Bethe Ansatz
equations \eref{eq:BAEpsi1}, \eref{eq:BAEksi1} and \eref{eq:BAEu1}
with the definitions \eref{eq:CDEdef} diagonalise the transfer 
matrix {\bf T} for arbitrary choice of the weights except for the
one constraint $r_3 t_1^2 t_5^2 = r_1 t_3^2 t_4^2$. It thus follows
that the point of maximum symmetry, where $r_i = r$ for $i\neq 2$
\footnote{We need to tune $r_2$ to compensate for the change in area
  induced by the transformation to the lattice.} and $t_i = t$, is
included in the spectrum obtained by this Ansatz. The eigenvalue of
{\bf T} in terms of solutions of the Bethe Ansatz equations is
given by \eref{eq:eigval}.   

\section{Integral equations}
To calculate the entropy, we put $t_i=1$, $r_2=\e^{\mu_1}$ and
$r_1=r_3=r_4=r_5=\e^{\mu_2}$. The chemical potentials $\mu_1$ and
$\mu_2$ have to be adjusted such that all configurations in the
original undeformed tiling model are weighted properly. The difference
between $\mu_1$ and $\mu_2$ compensates for the fact that the area of
the transformed tile r$_2$ is twice that of the other transformed
rectangles. If we want to weight the original rectangles in the random
tiling equally it follows that they must satisfy \cite{Li:1992}  
\begin{equation}
\mu_1-\mu_2 = N^{-1} \log \Lambda_{\rm max}. \label{eq:muval}
\end{equation}
One finds numerically that the maximum of the entropy of the tiling
model is at the point of maximum symmetry, i.e. where all
configurations are weighted equally, in agreement with the first random
tiling hypothesis \cite{Henley:1991}. According to \eref{eq:QCpartdens} this
is in the sector $n_1=m=\tau^{-2} N$, $n_2 = \tau^{-3}N$,
corresponding to tile fractions $n_{\rm rect} = \frac{5}{2}\tau^{-5}N$
and $n_{\rm tri} = 10 \tau^{-4} N$. 

\begin{figure}[h]
\centerline{
\begin{picture}(288,209)
\put(0,0){\epsffile{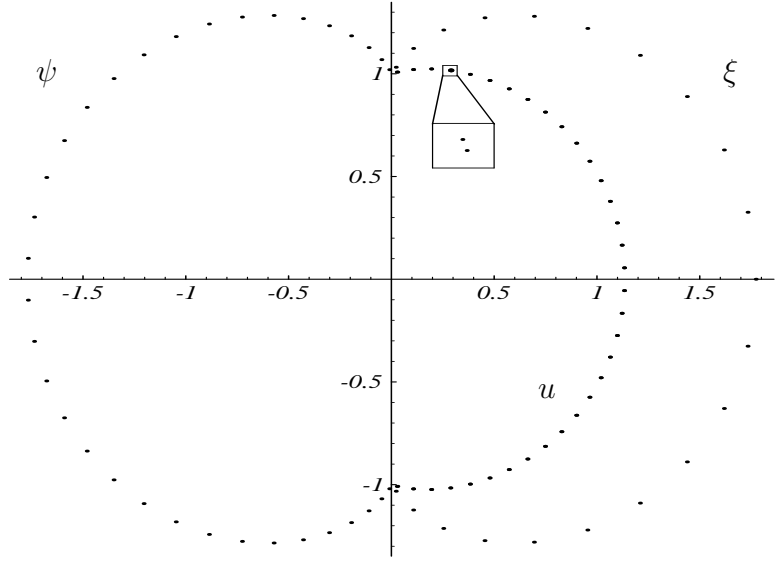}}
\put(10,180){$\psi$}
\put(200,60){$u$}
\put(270,180){$\xi$}
\end{picture}}
\caption{Distribution of roots for the largest eigenvalue ($N$ = 89,
  $n_1$ = m = 34, $n_2$ = 21). The left curve corresponds to the roots
  $\psi_i$ and the outer right curve corresponds to some subset
  $\{\xi_i\}$ of the roots $\tilde{\xi}_i$. The inner right curve
  actually consists of two curves corresponding to the roots $u_i$ and
  the subset of the roots $\tilde{\xi}_i$ complementary to
  $\{\xi_i\}$. Note that the inset is not on scale.} 
\label{fig:BAnumcurv}
\end{figure}
\begin{figure}[h]
\centerline{
\begin{picture}(253,253)
\put(0,0){\epsffile{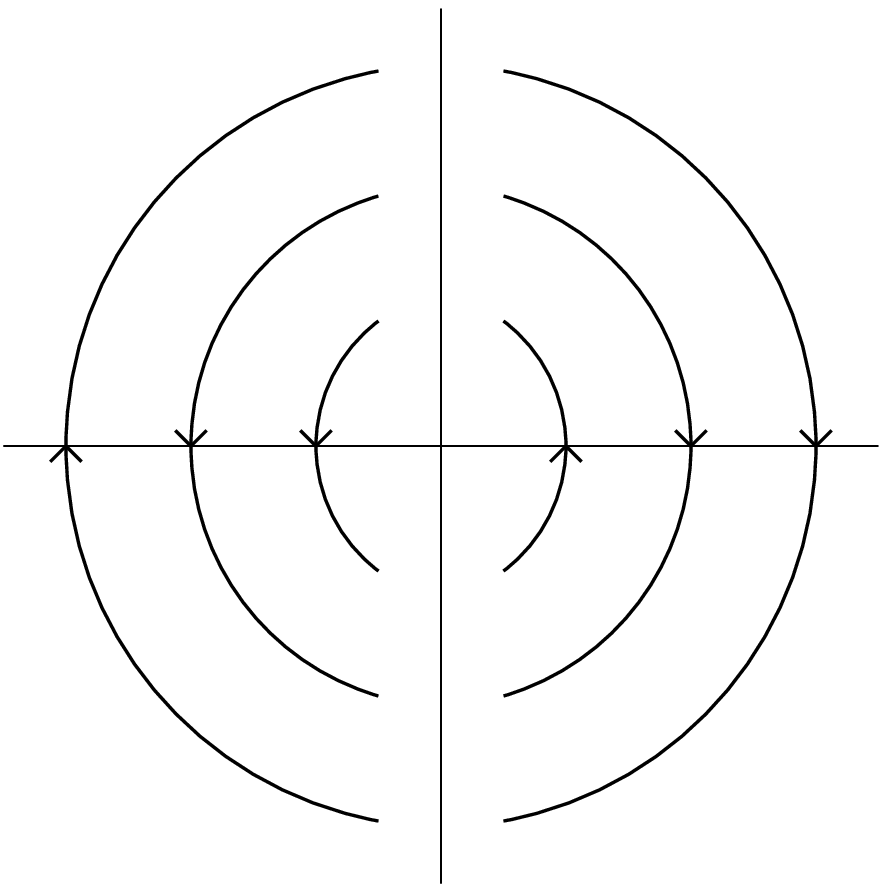}}
\put(100,10){$\psi_1$}\put(100,243){$\psi_m$}
\put(91,45){$-u_{n_1}^{-1}$}\put(91,208){$-u_1^{-1}$}
\put(91,80){$-\xi_{n_2}^{-1}$}\put(91,173){$-\xi_1^{-1}$}
\put(144,10){$\xi_{n_1}$}\put(144,243){$\xi_1$}
\put(144,45){$u_{n_2}$}\put(144,208){$u_1$}
\put(135,80){$-\psi_1^{-1}$}\put(135,173){$-\psi_m^{-1}$}
\end{picture}}
\caption{Schematic picture of the solution curves of the BA equations.}
\label{fig:BAcurves}
\end{figure}

It is also observed numerically that each of the roots $u_l$
approximates one of the roots $\tilde{\xi}_i$ in exponentially good
precision, see \fref{fig:BAnumcurv}. For reasons that will become
clear below, we define the roots $\xi_i$ now as the subset of the
roots $\tilde{\xi}_i$ {\em not} approximated by one of the $u_l$ and
we introduce the following notation    
\begin{equation}
s_\xi = \prod_{i=1}^{n_2} \xi_i^{1/2},\quad s_\psi = \prod_{j=1}^m 
(-\psi_j)^{1/2},\quad s_u = \prod_{l=1}^{n_1} u_l^{1/2}.
\end{equation}
By writing $\tilde{\xi}=u_l+\varepsilon_l$ for those roots
$\tilde{\xi}$ that are approximated by one of the $u_l$ and using the
above abbreviations, equation \eref{eq:BAEksi1} splits up in the
following two sets of equations,
\begin{eqnarray}
\xi_i^{(N+3n_1+n_2+m)/2} = (-)^{n_2-1} D s_\xi s_u s_\psi^{-1} 
\prod_{j=1}^m (\xi_i-\psi_j) \prod_{l=1}^{n_1}
(\xi_i-u_l)(\xi_i+u_l^{-1}),
\label{eq:BAEksi2} \\
(u_k+\varepsilon_k)^{(N+n_1+n_2+m)/2} = (-)^{n_1+n_2-1} D s_\xi s_u
s_\psi^{-1} \prod_{j=1}^m (u_k-\psi_j+\varepsilon_k) \times\nonumber\\
\hphantom{(u_k+\varepsilon_k)^{(N+n_1+n_2+m)/2} = }{}
\prod_{l=1}^{n_1} ( u_l-u_k-\varepsilon_k )
\left( 1+u_l^{-1}(u_k+\varepsilon_k )^{-1}\right).  
\label{eq:BAuexact}
\end{eqnarray}
Similarly, substituting $\tilde{\xi}=u+\varepsilon$ in \eref{eq:BAEu1}
results in 
\begin{eqnarray}
(-)^{n_1-1} E^{-1} u_l^{n_2} &=& 
\prod_{k=1}^{n_1} (u_l-u_k-\varepsilon_k) \left(
1+u_l^{-1}(u_k+\varepsilon_k)^{-1} \right)
\times\nonumber\\
&&\prod_{i=1}^{n_2} (u_l-\xi_i)(u_l+\xi_i^{-1}). \label{eq:BAEu2} 
\end{eqnarray}
In the thermodynamic limit all $\varepsilon_k$ vanish exponentially in
$N$. \Eref{eq:BAEu2} can then be used to remove the product over
the variables $u$ in \eref{eq:BAuexact} and we arrive at the following
equation which approximates the original BA equation \eref{eq:BAuexact}, 
\begin{eqnarray}
u_k^{(N+n_1-n_2+m)/2} &=& (-)^{n_1-1} D E^{-1} s_\xi s_u
s_\psi^{-1} \prod_{j=1}^m (u_k-\psi_j) \times\nonumber\\
&&\prod_{i=1}^{n_2} (\xi_i-u_k)^{-1} (u_k+\xi_i^{-1})^{-1} + 
{\cal O}(\e^{-N}),\label{eq:BAEuappr}
\end{eqnarray}
Consider now the BA equation \eref{eq:BAEpsi1} as a function of
$\psi_j$. Taking the logarithm on both sides of \eref{eq:BAEpsi1}
we define the function $F_\psi$ by
\begin{eqnarray}
F_\psi(z) &=& \log (-z)-\frac{2}{N+n_1+n_2+m} \left[ \sum_{i=1}^{n_2}
\log (\xi_i-z) + \sum_{l=1}^{n_1} \log (u_l-z) \right.\nonumber \\
&& \left.  \vphantom{\sum_{l=1}{n_1}} - \case{1}{4} N\mu_1 +
\case{1}{2}(n_1+n_2) \mu_2 - \log \frac{s_\xi s_u}{s_\psi} \right],
\label{eq:Fpsi}
\end{eqnarray}
so that $\Re F_\psi(\psi_j)=0$. Similar functions $F_\xi(z)$ and
$F_u(z)$ are defined by taking the logarithm of equations
\eref{eq:BAEksi2} and \eref{eq:BAEuappr} respectively. The BA
equations \eref{eq:BAEpsi1}, \eref{eq:BAEksi2} and \eref{eq:BAEuappr}
are then equivalent to  
\begin{equation}
\eqalign{
\case{1}{2} (N+n_1+n_2+m) F_\psi(\psi_k) &= 2\pi\i I_k,\\
\case{1}{2} (N+3n_1+n_2+m) F_\xi(\xi_k) &= 2\pi\i J_k,\\
\case{1}{2} (N+n_1-n_2+m) F_u(u_k) &= 2\pi\i K_k,}
\end{equation}
where $I_k$, $J_k$ and $K_k$ are either integers or half-integers.
From numerical calculations it follows that the numbers $I_k$, $J_k$
and $K_k$ for the solutions of the BA equations for the largest
eigenvalue are consecutive, more precisely  
\begin{equation}
\eqalign{
I_k &= \case{1}{2} (m+1-2k),\quad (k=1,\ldots,m)\\ 
J_k &= \case{1}{2}  (n_2+1-2k),\quad (k=1,\ldots,n_2) \\
K_k &=  \case{1}{2} (n_1+1-2k),\quad (k=1,\ldots,n_1) }
\label{eq:rootnu}
\end{equation}
We will assume that \eref{eq:rootnu} holds in the thermodynamic
limit. It is for this reason that we introduced the variables $\xi$
instead of $\tilde{\xi}$. The function analogous to \eref{eq:Fpsi}
defined from \eref{eq:BAEksi1} does not take consecutive multiples of
$2\pi \i$ when evaluated at the roots $\tilde{\xi}_k$.
It is clear from \eref{eq:rootnu} that the derivatives $f$ of the
functions $F$ are up to a factor precisely the densities of the Bethe
Ansatz roots. They allow us to transform the sums in the logarithmic
form of the BA equations into integrals. For brevity we define
\begin{equation}
\alpha=\frac{1+3Q_1+Q_2+Q_m}{1+Q},\quad 
\beta=\frac{1+Q_1-Q_2+Q_m}{1+Q},
\end{equation}
where $Q=Q_1+Q_2+Q_m$. Taking into account the root distribution
\eref{eq:rootnu} we thus arrive at the following integral equations
for the functions $f$,
\begin{eqnarray}
f_\psi (z) &=& \frac{1}{z} + \frac{\alpha}{2\pi\i} \int_{\xi_1}^{\xi_{n_2}} 
\frac{f_\xi(\xi)}{z-\xi} \d \xi + \frac{\beta}{2\pi\i} \int_{u_1}^{u_{n_1}} 
\frac{f_u(u)}{z-u} \d u, \label{eq:intfpsi}\\
f_\xi (z) &=& \frac{1}{z} + \frac{\beta}{2\pi\i\alpha} \left[ 
\int_{u_1}^{u_{n_1}} \frac{f_u(u)}{z-u} \d u +
\int_{-u_1^{-1}}^{-u_{n_1}^{-1}} \frac{u^{-2}f_u(-u^{-1})}{z-u} \d u
\right] \nonumber\\ 
&& {}+ \frac{1}{2\pi\i\alpha} \int_{\psi_1}^{\psi_m}
\frac{f_\psi(\psi)}{z-\psi} \d\psi ,\label{eq:intfxi}\\
f_u(z) &=& \frac{1}{z} - \frac{\alpha}{2\pi\i\beta} \left[
\int_{\xi_1}^{\xi_{n_2}} \frac{f_\xi(\xi)}{z-\xi} \d\xi + 
\int_{-\xi_1^{-1}}^{-\xi_{n_2}^{-1}} \frac{\xi^{-2}f_\xi(-\xi^{-1})}{z-\xi} 
\d\xi \right]\nonumber\\
&&{}+ \frac{1}{2\pi\i\beta} \int_{\psi_1}^{\psi_m} \frac{f_\psi(\psi)}{z-\psi}
\d\psi, \label{eq:intfu}
\end{eqnarray}
where the integrals are taken along the locus of the roots $\xi$,
$\psi$ and $u$ and $-\xi^{-1}$, $-\psi^{-1}$ and $-u^{-1}$. These
integration contours in the complex plane are schematically shown in
\fref{fig:BAcurves}.  

\section{Monodromy properties}
From the integral equations it is immediately seen that the integration
contours are cuts in the complex plane of the functions $f_\xi(z)$,
$f_\psi(z)$, $f_u(z)$, $z^{-2} f_\xi(-z^{-1})$, $z^{-2} f_\psi(-z^{-1})$ and  
$z^{-2} f_u(-z^{-1})$. Each of these functions has jumps across some
of the cuts whose magnitudes are given by linear combinations of the
above six functions. The analytic continuation across the cuts of
these functions is now determined by compensating for the jump and can
be written in terms of monodromy operators, one for each cut. The
analytic continuations across the different cuts of a function
$G(z)=\sum_{i=\xi,\psi,u} (a_i f_i(z) + b_i z^{-2} f_i (-z^{-1}))$ are
given by the following matrices which act on the vector ${\bf a} =
(a_\xi,a_\psi, a_u,b_{\xi},b_{\psi},b_{u})$,  
\begin{equation}
\eqalign{
\Gamma_\psi &= I -\alpha^{-1} E_{21} -\beta^{-1} E_{23},\\
\Gamma_{\psi^{-1}} &= I - \alpha^{-1} E_{54} - \beta^{-1} E_{56},\\
\Gamma_\xi &= I + \alpha E_{12} -\alpha\beta^{-1}
(E_{13}+E_{16}),\\
\Gamma_{\xi^{-1}} &= I + \alpha E_{45} - \alpha\beta^{-1}
(E_{43}+E_{46}),\\
\Gamma_u &= I + \beta E_{32} + \beta\alpha^{-1} (E_{31} + E_{34}),\\
\Gamma_{u^{-1}} &= I + \beta E_{65} + \beta\alpha^{-1} (E_{61}+E_{64}),}
\label{eq:monodr}
\end{equation}
where $I$ is the $6\times 6$ unit matrix and $E_{ij}$ is the matrix
with a 1 at the entry $ij$ and 0's everywhere else. As a result,
several linear combinations of the six functions, following from these
monodromy operations, correspond to  different sheets of the Riemann
surface  of one function. 

The continuation of a function $G(z)$ starting at and returning to the
origin across all of the six curves is given by the monodromy operator
$\Gamma=\Gamma_{\xi^{-1}} \Gamma_{u^{-1}} \Gamma_\psi \Gamma_\xi
\Gamma_u \Gamma_{\psi^{-1}}$ which has the property that $\Gamma^5 =
1$. This means that if the curves in \fref{fig:BAcurves} have 
the same endpoints, say $b$ for the common endpoint in the upper half
plane and $b^*$ for that in the lower half plane, that function is
single valued in terms of the variable $s$:
\begin{equation}
s(z) = \left( \frac{zb^{-1} -1}{1-z{b^*}^{-1}} \right)^{1/5},\quad
z(s) = b \frac {1+s^5}{1+b{b^*}^{-1} s^5}. \label{eq:sdef}
\end{equation}
When the BA-curves have the same endpoints, $\Gamma$ is the only non
trivial monodromy operator and the Riemann surface of $G(z)$
breaks up into infinitely many disconnected parts each 
with only five sheets. These are all mapped onto the plane by
\eref{eq:sdef}, from which it is easily seen that, apart from the
branch points $z=b$ and $z=b^*$, each point in the $z$-plane has five
images on the $s$-plane. As can be seen from \fref{fig:BAnumcurv}, for
the largest eigenvalue the curves do indeed have the same endpoints,
in particular they meet at $b=-b^*=\i$. The situation is rather
similar to what happens in the square triangle \cite{Kalugin:1994} and
the octagonal rectangle triangle tiling \cite{Gier:1996a}, where also
the BA curves close, but there the monodromy is of order 6 and 4
respectively. Although we do not understand the deeper reason for
this, it will become obvious in the sequel that the order of the
monodromy must relate to the symmetry of the tiling to produce the right
quadratic irrationalities. 

In the following we show that the closing of the curves enables us to
calculate the largest eigenvalue explicitly. Note that the six curves
can only have the same endpoints if $b=-b^{-1}=\i$, see
\fref{fig:BAcurves}, in contrast to the square triangle and the
octagonal rectangle triangle tiling, where the common endpoints need
not lie on the imaginary axis. We will, however, use the notation
$b=\i|b|\e^{\i\gamma}$ for the common endpoint of the curves, for it
may turn out in the future that for infinitesimal values of $\gamma$
still something can be said.

Aside from the cuts, the form $G\d z$ has only simple poles at $z=0$
and at  $z=\infty$. The poles and their residues of the single valued
function are thus given by the poles and residues on each of the
sheets. The functional form of $G(z)$ on other parts of the sheet and
on the other sheets is directly determined by the monodromy operators
\eref{eq:monodr}. In particular, the value of the residues can then be
read off from the integral equations \eref{eq:intfpsi},
\eref{eq:intfxi} and \eref{eq:intfu}. We choose $G(z)$ to be equal to
$f_\psi(z)$ near $z=0$ on the sheet with $s_1=\e^{\i \pi/5}$. Its form
near $z=0$ on the sheets with $s_{2k-1}=\e^{(2k-1) \i \pi/5}$ is
obtained by applying $\Gamma^k$. The functional forms of the function
$G(z)$ near $z=\infty$ on the sheets with $s_{2k} =
\e^{2\i(k\pi-\gamma)/5}$ are obtained by analytic continuation,
i.e. by applying $\Gamma_\xi \Gamma_u \Gamma_{\psi^{-1}}$ on each of
the functions corresponding to $s_{2k-1}$. The poles $s_n$ and
residues $R_n$ thus obtained of this function $G$ are shown in
\tref{tab:ResG}.    
\begin{table}[h]
\renewcommand{\arraystretch}{1.3}
\caption{Poles and residues of the form $G(z)\d z$. The two columns on
  the left list the poles of $G(z)\d z$ on the $z$- and on the
  $s$-plane respectively. The third column gives the functional form
  of $G(z)$ near those poles and the fourth column the value of the
  residues of these poles.}
\begin{indented}
\item[]
\begin{tabular}{llll}\br 
$z$ & $s_n$ & $G$ & $ R_n = {\rm Res}_{-1} (G\d z)$ \\ \mr
0 & $\e^{\pi\i/5}$ & $f_\psi(z)$ & 1\\
$\infty$ & $\e^{2\i(\pi-\gamma)/5}$ & $f_\psi(z)+\beta f_u(z)$ &
$\frac{-2(1+Q_2)}{1+Q}$\\ 
0 & $-\e^{-2\pi\i/5}$ & $\beta f_u(z)-\alpha z^{-2}f_\xi(-z^{-1})$ & 
$\frac{2}{1+Q}$\\
$\infty$ & $-\e^{-\i(\pi+2\gamma)/5}$ & $-\alpha z^{-2}f_\xi(-z^{-1})
+z^{-2}f_\psi(-z^{-1})$ & $\frac{-2Q_1}{1+Q}$ \\
0 & $-1$ & $z^{-2}f_\psi(-z^{-1})$ & $-\frac{1+Q_m-Q_1-Q_2}{1+Q}$\\
$\infty$ & $-\e^{\i(\pi-2\gamma)/5}$ & $z^{-2}f_\psi(-z^{-1})$ & 1\\
0 & $-\e^{2\i\pi/5}$ & $z^{-2}f_\psi(-z^{-1})+\beta z^{-2}f_u(-z^{-1})$
& $\frac{-2(1+Q_2)}{1+Q}$ \\
$\infty$ & $\e^{-2\i(\pi+\gamma)/5}$ & $\beta z^{-2}f_u(-z^{-1})
-\alpha f_\xi(z)$ & $\frac{2}{1+Q}$ \\
0 & $\e^{-\i\pi/5}$ & $-\alpha f_\xi(z)+f_\psi(z)$ & $\frac{-2Q_1}
{1+Q}$ \\
$\infty$ & $\e^{-2\i\gamma/5}$ & $f_\psi(z)$ & $-\frac{1+Q_m-Q_1-Q_2}
{1+Q}$ \\
\br
\end{tabular}
\end{indented}
\label{tab:ResG}
\end{table}

The form $G\d z$ is now uniquely determined by its poles and residues,
\begin{equation}
G\d z = \sum_{n=1}^{10} \frac{R_n}{s-s_n}\d s. \label{eq:1form}
\end{equation} 
Recall that we are working with approximated BA equations which are
valid for the largest eigenvalue, for which $Q_1=Q_m=\tau^{-2}$,
$Q_2=\tau^{-3}$ and thus $Q=1$. We do not know at the moment if the
approximation is valid for other sectors as well. The closing of the
three curves at $b=\i$ ($\gamma=0)$, however, is a fact that only
holds for the largest eigenvalue. The above arguments for calculating
$G\d z$ therefore are valid for this sector only, for which $G\d z$ is
given by 
\begin{equation}
G\d z = (s\e^{-2\i\pi/5}+s^{-1}\e^{2\i\pi/5}) \tau^{-1}  \frac{\d
  z}{z}. \label{eq:1formexpl}
\end{equation}
\section{Solution of definite integrals}
In this section we calculate some definite integrals of the form
\eref{eq:1formexpl}, resulting in an exact expression for the maximum of
the entropy. To remove singularities in the expressions to follow, we
introduce the following forms
\begin{equation}
g_n\d z = \frac{R_{2n-1}}{s-s_{2n-1}} \d s.
\end{equation}
In the $z$-plane we can calculate the following integrals using the
fact that $b=\i$ is a solution of the BA equations and that we have a precise 
control of the singularity at $z=0$ by using the mapping
\eref{eq:sdef} to the $s$-plane. 
\begin{eqnarray}
 J_{1,0} &=& \Re \left[ \int_b^0 (f_\psi(z)-g_1)\d z \right] =
\Re \lim_{s\rightarrow \e^{\i\pi/5}} \left[ F_\psi(z(s)) - \log
(s-\e^{\i\pi/5}) \right] \nonumber\\
 &=& \log (5/2) -N^{-1}\log\Lambda +\mu_1 \frac{1}{2} -\mu_2 \tau^{-2},\\
 J_{2,0} &=& \Re \left[ \int_b^0 (\beta f_u(z) - \alpha z^{-2}
f_\xi(-z^{-1}) -g_2)\d z \right] \nonumber\\
 &=& \log(5/2) -2N^{-1}\log s_\psi +\mu_1 \frac{1}{2}\tau^{-1} -
\mu_2\tau^{-2},\\ 
 J_{3,0} &=& \Re\left[ \int_b^0 (z^{-2}f_\psi(-z^{-1})-g_3)\d z \right]
\nonumber\\
 &=& -\tau^{-2}\log(5/2) +N^{-1}\log\Lambda -2N^{-1}\log s_\psi
-\mu_2\tau^{-3},\\ 
 J_{4,0} &=& \Re\left[ \int_b^0 (z^{-2}f_\psi(-z^{-1})+\beta
z^{-2}f_u(-z^{-1}) -g_4)\d z \right] \nonumber\\
 &=& -2\tau^{-1}\log(5/2) +\mu_1\tau^{-1}-\mu_2 2\tau^{-2},\\
 J_{5,0} &=& \Re\left[ \int_b^0 (-\alpha f_\xi(z) +f_\psi(z))-g_5)\d z
\right] \nonumber\\
 &=& -\tau^{-2}\log(5/2) +\mu_1 \frac{1}{2}\tau^{-2}-\mu_2\tau^{-3}
= \frac{1}{2\tau} J_{4,0}.
\end{eqnarray}
The same integrals on the $s$-plane restricted to the symmetric
point are, using \eref{eq:1form},
\begin{equation}
\eqalign{
J_{1,0} &= \sum_{n=2}^{10} R_n \log |\e^{\i\pi/5}-s_n| = \tau\log\tau
- \log \sin (2\pi/5),\\ 
J_{2,0} &= \tau\log\tau - \log \sin (2\pi/5),\\
J_{3,0} &= -\tau^{-1}\log\tau +\tau^{-2} \log\sin(2\pi/5),\\
J_{4,0} &= -2\log\tau +2\tau^{-1} \log\sin(2\pi/5),\\
J_{5,0} &= -\tau^{-1}\log\tau +\tau^{-2} \log\sin(2\pi/5).}
\end{equation}
Equating both sets of integrals gives the following solution
\begin{eqnarray}
4\tau^{-1} s_\psi &=& \mu_1 - (1+ \tau^{-3})\mu_2 = N^{-1}\log\Lambda 
- \tau^{-3}\mu_2 \nonumber\\
&=& \frac{1}{2} \left( \log \frac{5^5}{4^4} - 2\sqrt{5} \log \tau \right).
\end{eqnarray} 
Note that this solution precisely corresponds to the maximum of the
entropy for the original tiling model according to \eref{eq:muval}.
The total number of rectangles per site, $Q_{\rm rect} = n_{\rm
  rect}/N$, on the symmetric point is $\case{5}{2}\tau^{-5}$, so that the entropy per site is given by 
\begin{equation}
\sigma_N = N^{-1}\log\Lambda - \case{1}{2}\tau^{-5}(\mu_1+4\mu_2) 
= \case{5}{2}\tau^{-2} (N^{-1}\log\Lambda - \tau^{-3}\mu_2).
\end{equation}
The number of vertices per site $n_{\rm v}$ is given by
\begin{equation}
n_{\rm v} = \case{1}{2} Q_{\rm tri} + Q_{\rm rect} =
\case{5}{2} \tau^{-4} (2+\tau^{-1}) = \case{5}{2}\tau^{-2}.
\end{equation} 
Thus, the entropy per vertex $\sigma_{\rm v}$ is finally given by
\begin{equation}
\sigma_{\rm v} = N^{-1} \log\Lambda - \tau^{-3}\mu_2 = \frac{1}{2} 
\left( \log \frac{5^5}{4^4} - 2\sqrt{5} \log \tau \right).
\end{equation} 
\section{Conclusion}
In this paper we showed that a decagonal random tiling model of
rectangles and triangles is solvable using the Bethe Ansatz
technique. We derived the Bethe Ansatz equations that diagonalise the
transfer matrix for this model. These equations contain all
information about the model and they in principle present a huge
reduction of computational problems concerning the system size. For
this tiling model however, some of the roots of the BA equations
almost coincide, which makes it difficult to extract high precision
data. On the other hand, it enabled us to write down approximate BA
equations which are exact at the symmetric point in the thermodynamic
limit. Using these equations we were able to find an exact expression
for the maximum of the entropy. The validity of the approximation
outside the symmetric point still has to be further investigated. We
hope to be able to find analytic expressions for the phason elastic
constants as well, although in contrast to the solutions of the
dodecagonal square-triangle \cite{Widom:1993,Kalugin:1994} and the
octagonal rectangle triangle \cite{Gier:1996a,Gier:1997b} it is not
apparent if our solution can be extended off the symmetric
point. Recently, Kalugin \cite{Kalugin:1997} showed for the square
triangle model that also critical exponents may be calculated exactly
from the BA equations.  

\ack
We thank M Martins for discussions and A Verberkmoes for a reading of
the manuscript. This research was supported by `Stichting Fundamenteel
Onderzoek der Materie' which is financially supported by the Dutch
foundation for scientific research NWO.

\end{document}